\documentclass[preprint,nofootinbib,showpacs,aps,prc]{revtex4-1}

\usepackage{graphicx}
\usepackage{amstext}
\usepackage{amssymb}
\usepackage{amsmath}
\begin{document}
\title{Correlation femtoscopy of small systems}

\author{Yu.M. Sinyukov}
\author{V.M. Shapoval}
\affiliation{Bogolyubov Institute for Theoretical Physics, Metrolohichna str. 14b, 03680 Kiev, Ukraine}

\begin{abstract}
The basic principles of the correlation femtoscopy, including its correspondence to the Hanbury Brown and Twiss intensity interferometry,  are re-examined. The main subject of the paper is an analysis of the correlation femtoscopy  when the source size is as small as the order of the
uncertainty limit. It is  about 1 fm for the current high energy experiments. Then the standard femtoscopy model of  random sources is inapplicable. The uncertainty principle leads to the partial  indistinguishability and coherence of closely located emitters that affect the observed femtoscopy scales. In thermal systems the role of corresponding coherent length is taken by the thermal de Broglie wavelength that also defines the size of a single emitter. The formalism of partially coherent phases in the amplitudes of closely located individual emitters is used for the quantitative analysis. The general approach is illustrated analytically for the case of the Gaussian approximation for emitting sources. A reduction of the interferometry radii and a suppression of the Bose-Einstein correlation functions for small sources due to the uncertainty principle are found. There is a positive correlation between the source size and the intercept of the correlation function. The peculiarities of the non-femtoscopic correlations caused by minijets and fluctuations of the initial states of the systems formed in $pp$ and $e^+e^-$  collisions are also analyzed. The factorization property for the contributions of femtoscopic and non-femtoscopic correlations into complete correlation function is observed in numerical calculations in a wide range of the model parameters.
\end{abstract}

 \maketitle


Corresponding author: {\small \textit{Yu.M. Sinyukov, Bogolyubov Institute
for Theoretical Physics, Kiev 03680, Metrolohichna 14b, Ukraine. E-mail:
sinyukov@bitp.kiev.ua}}

\section{Introduction}
The correlation femtoscopy, or intensity interferometry method, is the
direct tool to measure the spatial and temporal scales of extremely
small and short-lived systems created in particle and nuclear collisions with accuracy
of $10^{-15}$ m and $10^{-23}$ sec, respectively. The method \cite{Gold,Kopylov, Coc}, is grounded on the
Bose-Einstein (BE) or Fermi-Dirac (FD) symmetric properties of the quantum states. It has a deep analogy with the intensity interferometry
telescope that was proposed by Hanbury Brown and Twiss for measurements of
angular sizes of remote stars \cite{HBT}. In distinction on standard telescopic
and microscopic techniques based on the registration of intensities of light
or particles, e.g., electrons, coming from (or through) the object, this method deals
with the correlation between intensities of the source radiation registered
by two (many) spatially separated parts of devices, such as telescopes, reflectors,
particle detectors, etc. In fact, it measures the correlations between
numbers of emitted identical particles detected in separated parts of the detector.

The femtoscopic space-time structure of the systems is typically represented in terms of the interferometry radii. They are
result of the Gaussian fit of the correlation function defined as a ratio of
the two- (identical) particle spectra to the product of the single-particle
ones. In the pioneer papers \cite{Kopylov, Coc} the measured interferometry radii were
interpreted as the geometrical sizes of the systems. Later on it was found
\cite{Pratt,MakSin, Hama} that for typical systems formed in experiments with heavy ions, the
above geometrical interpretation needs to be generalized.
The treatment of the interferometry radii as the homogeneity lengths \cite{Sin, AkkSin} in
the systems and the crucial suggestion for femtoscopy scanning of the source
radiation in different momentum bins bring the possibility to analyze different parts of the source and explain the behavior of the interferometry radii.
In addition, the practical method how to use the final state interactions (FSI) and effects of long-lived resonances to extract the BE correlations in relatively large systems created in heavy ion collisions has been proposed \cite{FSI}.

The other challenge, which is still actual, concerns the femtoscopy analysis of relatively  small systems created in particle interactions such as $pp$ and $e^+e^-$, where the observed femtoscopic scales are approximately 1 fm or smaller \cite{Kittel}. Typically, the suppression of the correlation function is fairly large in these processes. Here we will analyze the femtoscopy of such small systems accounting for the uncertainty principle,  coherence of the radiation from spatially very closely set emitters  and non-femtoscopic (non BE, FD and FSI) correlations. The latter appear due to the energy-momentum conservation law and  incoherent contributions to the two- and single-particle spectra induced by particle clusterization in momentum space and fluctuations of initial conditions of the collision processes. The detailed analysis  of these theoretical problems can help to provide the correct femtoscopy study of the small systems.

\section{The basic ideas of the Intensity Interferometry Telescope}

The intensity interferometry method for the measurement of the stars' angular sizes was proposed
and realized first by Hanbury Brown and Twiss \cite{HBT} at the
end of the 1950s.  The electromagnetic radiation from the star is the mixture of different, almost monochromatic wave trains, which are mutually incoherent at the moment of radiation. To see the principal aspects, let us consider the emission from the different sites  of a radiating object. If the stellar object is close to the observer, like our Sun, then one can easily select, say, the opposite sites (edges) of it. If the two
telescopes are directed to those different sites and one measures the correlations between photon numbers coming to each of the two telescopes, then the waves  from
different sites of the source do not mix in the telescopes, and the correlations between them are absent --- the signals coming to the two telescope reflectors are mutually incoherent. If, however, the stellar object --- let us consider now the double star system --- is very remote, so that it is
impossible to select only one of the two stars by a telescope, the light from
both stars will come to each of the two reflectors and become mutually (partially) coherent.  For simplicity let us imagine that at some moment in time the telescopes register only one wave train from each star, and both these trains are equally polarized and follow each other continuously. {\it Ipso facto} we ignore the real problems of the intensity interferometry method --- how to extract the signal from the noise --- but preserve the principal point of this method.

\subsection{The basic formalism}
One can decompose the electric field strength into positive and negative
frequency parts, $\mathbf{E}=\mathbf{E}^{(+)}(\mathbf{x},t)+\mathbf{E}^{(-)}(\mathbf{x},t)$, $\mathbf{E}^{(-)}(\mathbf{x},t)=\left( E^{(+)}(\mathbf{x},t)\right) ^{\dagger }$. The ideal photon counter reacts just to the product $\mathbf{E}^{(-)}(\mathbf{x},t)\mathbf{E}^{(+)}(\mathbf{x},t)=|\mathbf{E}^{(+)}(\mathbf{x},t)|^2$  \cite{Glauber}. Far from the stellar object the light is described well by the plane waves, so that

\begin{equation}
\mathbf{E}^{(+)}(\mathbf{x},t)=Ae^{i(\mathbf{k}_{1}\cdot \mathbf{x}-\omega_1
t)}+Be^{i(\mathbf{k}_{2}\cdot \mathbf{x}-\omega_2 t)}  \label{electrowave}.
\end{equation}
The complex amplitudes $A$ and $B$ have stochastic independent phases: $A=|A(t)|e^{i\varphi_{A}(t,{\bf x})}$,
$B=|B(t)|e^{i\varphi_{B}(t,{\bf x})}$, which are roughly constants during coherence time $\tau_{coh}$ of the wave train, e.g., $\tau_{coh} \approx 10^{-8}$ sec, so that being averaged over a period of time $T \gg \tau_{coh}$, the amplitudes and their product become zero: $\langle A\rangle =0,$ $\langle B\rangle =0$, $\left\langle AB\right\rangle=0$.
The intensity of light that is proportional to the number of photons registered by the telescope/reflector and photo-multipliers at point $\mathbf{x}$ is

\begin{eqnarray}
 \left\langle I_{\mathbf{k}_{1},\mathbf{k}_{2}}(x)\right\rangle  &=&  \left\langle \mathbf{E}
^{2}\right\rangle =\left\langle \mathbf{E}^{(-)}(\mathbf{x},t)\mathbf{E}
^{(+)}(\mathbf{x},t)\right\rangle =\left\langle |\mathbf{E}^{(+)}(\mathbf{x},t)|^{2}\right\rangle \nonumber \\
&& \label{in} \\
&=&\left\langle |A|^{2}+|B|^{2}+AB^{\ast }e^{i(\mathbf{(k}_{1}\mathbf{-k}_{2}
\mathbf{)}\cdot \mathbf{x}-(\omega_1-\omega_2)t)}+A^{\ast }Be^{-i(\mathbf{(k}_{1}\mathbf{-k}_{2}\mathbf{)}\cdot \mathbf{x}-(\omega_1-\omega_2)t)}\right\rangle \nonumber \\
&=& |A|^{2}+|B|^{2},
\nonumber
\end{eqnarray}
where we supposed that $|A(t)|, |B(t)|$ are almost independent on $t$.

The statistically averaged intensity registered by one of the
telescopes contains the information only about the (averaged) squared modulus of the amplitudes. However, the correlation of intensities $C$, defined as the ratio of the averaged product of intensities
registered at the two space-time points $({\bf x_1},t_1)$ and $({\bf x_2},t_2)$ to the product of averaged intensities registered at these points,  depends already on the differences of momenta and energies of the light quanta. To simplify notation, we put $\left| A\right| =\left| B\right| $ and get
\begin{eqnarray}
C &=& \frac{\left\langle I_{\mathbf{k}_{1},\mathbf{k}_{2}}(\mathbf{x}_{1})I_{\mathbf{k}_{1},\mathbf{k}_{2}}(\mathbf{x}_{2})\right\rangle}{
\left\langle I_{\mathbf{k}_{1},\mathbf{k}_{2}}(\mathbf{x}_{1})\right\rangle \left\langle I_{\mathbf{k}_{1},\mathbf{k}_{2}}(\mathbf{x}_{2})\right\rangle}=\frac{1}{4|A|^{4}}\left\langle |E^{(+)}(\mathbf{x_{1}},t)|^{2}|E^{(+)}(\mathbf{x_{2}},t)|^{2}\right\rangle \nonumber  \\
\nonumber \\
&=& 1+\frac{1}{2}\cos [(\mathbf{k}_{1}\mathbf{-k}_{2})(\mathbf{x}_{1}-\mathbf{x}_{2})-(\omega_1-\omega_2)(t_1-t_2)]\label{intcorr}\\
& \approx &1+\frac{1}{2}\cos [\theta \left| \mathbf{k}\right|d+((x_{L1}-x_{L2})/c-t_1+t_2)(\omega_1-\omega_2)],\nonumber
\end{eqnarray}
where one took into account that $\langle |A|^{2}A^{\ast }B=0\rangle $, etc. Here the $\theta =D/L$ is the angular size of the double star system with ``transverse'' distance $D$ between stars in the plane perpendicular to the direction to the system, $L$ is the distance to the system, $d$ is the ``transverse'' distance between two telescopes, $\left| \mathbf{k}\right|$ is the mean detected wave number, $x_{iL}$ is the ``longitudinal'', directed to the system,  coordinate of the {\it i}-telescope and $t_1-t_2\equiv \Delta$ is the signal delay between points ${\bf {x}_1}$ and ${\bf{x}_2}$. It is worth noting that the time resolution $\tau_{res}$ in the method has to be smaller than the coherence time, $\tau_{res} < \tau_{coh}$, in order to provide correlation measurement of photon numbers during the mutual coherence time $\tau_{coh}$ of the electromagnetic waves in points ${\bf {x}_1}$ and ${\bf{x}_2}$. For a stationary process the averaging over a large period of time $T$ plays the role of the averaging over the ensemble of events with duration time $\tau_{coh}$. Also note that the condition of the validity of formula (\ref{intcorr}) is
\begin{equation}
\frac{x_{2L}- x_{1L}}{c} - (t_2-t_1)< \tau_{coh}.
\label{condition}
\end{equation}
Otherwise, $\langle
A({\bf x_1},t_1)A({\bf x_2},t_2)\rangle =0, \langle B({\bf x_1},t_1)B({\bf x_2},t_2)\rangle =0$, and the correlations disappear, $C=1$, because of the mutual incoherence of waves coming to the two telescopes.

\subsection{The nature of the HBT effect}
 The formula (\ref{intcorr}) demonstrates the principle of measurement the differences in momenta (and energies) of photons radiated by remote stars by measuring the correlation function depending on distances between telescopes/reflectors and time delay. In fact, the momentum difference in the transverse plane  is  connected with the angular size of a stellar object. As for the difference in the energy of the photons, it is possible, in principle, to measure it with the restriction given by Eq. (\ref{condition}), and this difference would  be associated with different temperatures of the two stars, if such a situation could take place. However, there is no direct connection of the difference in the energy of radiated photons with the time and space scales of the stellar system, and, therefore, only the angular size of this object can be  extracted in this way. The latter is the basic application of the intensity interferometry telescope, and the method  was used to measure the angular sizes of single remote stars
\footnote{In this case one deals not with two emitters, but with many emitters $N$ at the whole stellar disk, and formula (\ref{intcorr}) is modified at $t_1=t_2$ as follows: $C-1 \propto \frac{1}{N^2}\sum_{j\neq i=1}^N \exp{[i({\bf k}_{Ti}-{\bf k}_{Tj})({\bf x}_{T1}-{\bf x}_{T2})]}\stackrel{N\rightarrow \infty}{\rightarrow}(\frac{2J_1(\theta|{\bf k}|d)}{\theta|{\bf k}|d})^2$. Determining the distance corresponding to the first zero of the correlation function in this representation Hanbury Brown and Twiss have measured the angular size of Sirius and some other stars.}.

 Despite the classical description of the Hanbury Brown and Twiss (HBT) method, the detailed analysis of the measurements accounting for the principle of photon registration pointed to the quantum nature of the effect \cite{Glauber}. The relationship between classical and quantum descriptions of the electromagnetic waves has been established through the formalism of coherent states \cite{Glauber}. Appealing to the quantum nature of the electromagnetic fields,  one can say that the method of the measurement of the star angular size  is based on the positive correlations between numbers of photons registered  in  two close \textit{space-time} points because of Bose-Einstein statistics for these quanta.

If a double star system is so far from the detectors that the latter can register only a few photons per $\tau_{res}$, then one has to use the amplitudes of registration at points $x_1$ and $x_2$ of the two photons emitted with mutually random phases from the two stars $a$ and $b$: $A(x_1)=\left\langle a|x_1\right\rangle + \left\langle b|x_1\right\rangle$ and $A(x_2)=\left\langle a|x_2\right\rangle + \left\langle b|x_2\right\rangle$. Then taking the ratio of the averaged  modulus squared of symmetric (over photons permutation) two-photon amplitude $A(x_1,x_2)=A(x_1)A(x_2)$ to the product of the averaged single-photon amplitudes one can get again the correlation structure (\ref{intcorr}). Note that describing a registration of the two neutrinos from the star, one should use the antisymmetrized amplitude $A(x_1,x_2)= \left\langle a|x_1\right\rangle\left\langle b|x_2\right\rangle - \left\langle a|x_2\right\rangle\left\langle b|x_1\right\rangle$.

\section{The basic ideas of the correlation interferometry
}
\bigskip In particle physics the positive correlations between numbers of
identical pions with close momenta emitted from an interaction
region in proton-antiproton annihilations  were found in 1960 by Goldhaber \textit{et al.}
(GGLP effect) \cite{Gold}. It was understood that the nature of the effect lies in
quantum statistics for identical particles demanding the symmetrization of
bosonic wave function. Later on, based on this fundamental principle,  Kopylov and Podgoretsky \cite{Kopylov} developed the method of the pion
interferometry microscope, or correlation femtoscopy/interferometry. They found an analogy  between the correlation interferometry and Hanbury Brown and Twiss (HBT) stellar intensity interferometry \cite{HBT}. As one can see below, the basic mathematical structure of the two methods can be presented in identical form. However, at the HBT measurements, the interference of intensities/photon numbers  happen near the detector (telescope pair), where the correlations form, while in the femtoscopy the quantum statistical (QS) correlations arise in the emitting object. The HBT is based on the analysis of particle correlations as they depend on the space-time separation of the telescopes/reflectors, while the correlation interferometry deals with  dependence of the correlations on the particles' momenta differences. Correspondingly, the measurands are different. The HBT method measures differences of momenta and energies of photons radiated by stellar objects, and the only angular sizes  associated with observed momenta difference, but not spatiotemporal scales can be extracted \footnote {If the distance to the double star system is known, the only transverse  (to the direction of the stellar system)  projection of the distance between the stars is possible to restore from the angular size.}. In contrast, the correlation femtoscopy measures the space and time separation of the emission points, and so extracts all the sizes and three-dimensional geometrical shape of the source, as well as a duration of the emission. In this sense the term ``HBT radii'' that one often uses to present the femtoscopy measurements is not quite adequate, as is stressed in Ref. \cite{Ledn1}. 

\subsection{Standard approach}
 The basic ideas of the correlation femtoscopy are described in many publications, e.g., Ref. \cite{Gyulassy}. We reproduce them here with some important remarks.  Let us suppose that two identical bosons (e.g., pions) are emitted
from the two space-time points, $x_1=(t_1,\mathbf{x}_{1})$
and $x_2=(t_2,\mathbf{x}_{2})$, and then {\it propagate freely}. The wave function of a single particle at the initial time $t_i$  in the
configuration representation is\footnote{We use notation ($\tau,{\bf r}$) for current Minkowski coordinates to escape confusion with emission points $t,{\bf x}$.} $\delta^3({\bf r-x_i})$. At some time $\tau$   in the  \emph{momentum} representation with $p=(p^0=E,{\bf p})$, it is $\psi_{x_i}(p,\tau)= \frac{1}{(2\pi)^{3/2}}e^{-iE\tau}e^{ipx_i}$, where $p\,x_i=p^0t_i-{\bf p}{\bf x}_i$. Here and below we use dimensionless units: $\hbar=c=1$.

In the momentum representation ($p_{1},p_{2}$) the two-boson wave function is symmetrized and has the form
\begin{equation}
\psi _{x_{1},x_{2}}(p_{1},p_{2};\tau)=\frac{1}{\sqrt{2}(2\pi)^3}\left[ e^{ip_1x_1
}e^{ip_{2}x_{2}}+e^{ip_{2}x_1}e^{ip_1x_{2}}\right]e^{-i(E_1+E_2)\tau}.  \label{psi1}
\end{equation}
Then the probability to find the two pions with momenta $p_{1}$, $p_{2}$ is expressed through the scalar product of 4-momentum and 4-coordinate differences of the two-pion emission:

\begin{equation}
W_{x_{1},x_{2}}(p_{1},p_{2})=\left| \psi _{x_{1},x_{2}}(p_{1},p_{2};\tau)\right|
^{2}\propto  1+\cos \left[ (p_{1}-p_{2})\cdot (x_{1}-x_{2})\right].
\label{W(p1p2)}
\end{equation}

Comparing Eq.~(\ref{W(p1p2)}) with result (\ref{intcorr}), related to the interferometry telescope, one can see that
they differ from each other by a factor of $1/2$ before the cosine. The attentive reader can
notice the difference between the two cases. In the case of  emission from a double star system
the correlation of intensities accounts for all the possibilities: this is the correlation between intensities of the wave trains coming from  different stars as well as from the same star.
In contrast, when formula (\ref{W(p1p2)}) is derived, it
is supposed that if one boson (with momentum $\mathbf{p}_{1}$) is emitted  from
the point $\mathbf{x}_{1}$, then another boson (with momentum $\mathbf{p}_{2}$) is emitted from different point $\mathbf{x}_{2}$ and vice versa, but
the possibility when the two bosons are emitted from the same point, $\mathbf{x}_{1}$ or $\mathbf{x}_{2}$, is excluded. In the case of \textit{independent } particles' radiation all the possibilities have to be taken into account. For such a case one usually demonstrates the idea of the correlation femtoscopy method by means of factorization of the two-particle normalized emission function, $\rho(x_1,x_2)=\rho (x_1)\rho (x_2)$ and integrates the two-point probability (\ref{W(p1p2)}) over the space-time region. Then for the correlation function $C(p_1,p_2)$ one has (we ignore here possible correlations between coordinates and momenta of the emitted particles)
\begin{eqnarray}
C(p_1,p_2) &=& \frac{W(p_{1},p_{2})}{W(p_{1})W(p_{2})} =\frac{1}{W(p_{1})W(p_{2})} \int d^{4}x_{1}d^{4}x_{2}\rho (x_{1})\rho(x_{2})W_{x_{1},x_{2}}(p_{1},p_{2}) \nonumber  \\
&=&1+\frac{1}{W(p_{1})W(p_{2})}\left| \int d^{4}x\rho (x)e^{iq\cdot x}\right|^{2},
\label{femtobasic}
\end{eqnarray}
where $q=p_{1}-p_{2}$. So, the
probability to find the particles with momenta $p_{1}$, $p_{2}$ at very large times  $\tau\rightarrow \infty$ is expressed through the Fourier image of the emission function. This is the typical basis of the correlation interferometry method, allowing one to analyze the size and shape of small systems. If $\rho$ is the Gaussian-like
emission probability that in some reference frame has the form $\rho (x)\propto\exp \left[-\sum\limits_{i=1}^3
\frac{x_{i}^{2}}{2R_{i}^{2}}\right]\delta(t-t_0)$, then (\ref{femtobasic}) reads as
\begin{equation}
C(p_1,p_2)= 1+\exp\left[-\sum\limits_{i=1}^3
q_i^2R_{i}^{2}\right].
\label{Gauss}
\end{equation}
For example, the typical resolving width of the
correlation function $\overline{q}=50$ MeV corresponds to the size $4\cdot
10^{-15}$ m $=$ $4$ Fm (Fermi) $=$ $4$ fm (femtometer). The origination of
the method's name --- the correlation femtoscopy --- is obvious from
such estimates.

\subsection{Correlation femtoscopy formalism under microscope \label{Micro}}
The problem, however, appears when we apply the basic formula (\ref{femtobasic}) to the system with small  number of emitters. For example, if there are only two different emitting points: $\rho(x) = \frac{1}{2}(\delta^4(x-x_1)+\delta^4(x-x_2))$, then from (\ref{femtobasic}) follows ($\rho_i =1/2$)
\begin{eqnarray}
W_{x_{1},x_{2}}(p_{1},p_{2}) &\propto& \sum\limits_{i_1,i_2=1,2}
\rho_{i_1}\rho_{i_2}(1+\cos \left[(p_{1}-p_{2})(x_{i_1}-x_{i_2})\right])= \nonumber \\
&=&1+\cos^2\left[\frac{1}{2}(p_1-p_2)(x_1-x_2)\right].
\label{wrong}
\end{eqnarray}
The result is incorrect as we shall show.

To analyze the situation in detail, let us consider single-particle radiation from the two points. 
If the emission from the point $x_{1} $ is associated with the quantum state that is {\it distinguishable} and {\it independent} from the state corresponding to emission
from the point $x_{2}$, then the two orthogonal states~$A_{i}$
are allowed to be realized with the following known probabilities~$\rho _{i}$:
\begin{equation}
\rho_{1} \mbox{:\ \ \ }A_{1}(p)=e^{ipx_{1}}e^{-iE\tau}\mbox{\ \ \ } \mbox{and}\mbox{ \ \ \ }\rho_{2}\mbox{:\ \ \ }
A_{2}(p)=e^{ipx_{2}}e^{-iE\tau}\mbox{; \ \ \ }\rho_1+\rho_{2}=1.
\label{single}
\end{equation}
Here and below we omit multiplier $(2\pi)^{-3/2}$ since these factors cancel in the correlation function (see below). Note that $W_{x_{1},x_{2}}(p)=\rho _{1}A_{1}(p)A_{1}^*(p)+\rho_{2}A_{2}(p)A_{2}^*(p)=1$.
Since two identical bosons are emitted {\it independently} from the two points, there are the
three different final states (amplitudes) $A_{i_1i_2}$ that are distinguishable and realized with the
probabilities $\rho _{i_1i_2}$:
\begin{eqnarray}
\rho_{11}\mbox{: \ }
A_{11}(p_{1},p_{2})=e^{ip_{1}x_{1}}e^{ip_{2}x_{1}}e^{-i(E_1+E_2)\tau}\mbox{; \ }\rho_{22}\mbox{: \ }A_{22}(p_{1},p_{2})=e^{ip_{1}x_{2}}e^{ip_{2}x_{2}}e^{-i(E_1+E_2)\tau}\mbox{; \ }  \nonumber \\
&&  \label{double} \\
\rho_{12}\mbox{: \ \ }A_{12}(p_{1},p_{2})=\frac{1}{\sqrt{2}}
\left( e^{ip_{1}x_{1}}e^{ip_{2}x_{2}}+e^{ip_{1}x_{2}}e^{p_{2}x_{1}}\right)e^{-i(E_1+E_2)\tau}
\mbox{ ; \ }\rho_{11}+\rho_{22}+\rho_{12}=1.  \nonumber
\end{eqnarray}
If $\rho_{1}=\rho_{2}=1/2$, then $\rho _{ii}=\rho _{i}^{2}=1/4$ and $\rho _{12}=1/2$. As a result, the probability of finding the
two particles with momenta $p_{1},p_{2}$ that are emitted independently from the two
points $x_{1}$ and $x_{2}$ is
\begin{equation}
C(p_1,p_2)= \frac{W_{x_{1},x_{2}}(p_{1},p_{2})}{W_{x_{1},x_{2}}(p_{1})W_{x_{1},x_{2}}(p_{2})}=\sum\limits_{i_1\leqslant i_2=1,2}\rho
_{i_1i_2}\left| A_{i_1i_2}\right| ^{2}=1+\frac{1}{2}\cos \left[(p_1-p_2)(x_{1}-x_{2})\right].  \label{correct}
\end{equation}
The last result coincides with Eq.(\ref{intcorr}) for stellar
intensity interferometry telescope. The result (\ref{correct})
takes into account that the emission of both quanta from the same  point (say,~$x_{1}$) brings no interference, while the wrong account for the interference effect leads to the results like (\ref{wrong}). Only in the case of a very large number of emitters, when contributions to the correlation function from one-point two-particle radiations are relatively small, can one use the idealized continuous limit \,(\ref{femtobasic}). Therefore the simplest ''derivation'' of
the intensity interferometry method (\ref{femtobasic}) has
to be corrected to exclude the double accounting in the correlation function for the contribution associated with the particle pairs emitted from very close points. This correction is significant if the number of {\it independent} emitters is not large. For small systems with not flat momentum spectrum, there is essentially finite number of independent incoherent emitters because of the uncertainty principle. We shall consider these effects in detail in the next subsections.

If the emission from points $x_1$ and $x_2$ is not independent at all (full coherence), then one has to use not the  probabilities $\rho _{1},\rho _{2}$, but the pure state only to account for the interference between the different states:
$A_{x_1,x_2}(p)=A_{1}(p)+A_{2}(p)$.
The  amplitude of the two identical bosons symmetrized over $p_1,p_2$ in this case is: $A_{x_1,x_2}(p_1,p_2)=A_{x_1,x_2}(p_1)A_{x_1,x_2}(p_2)$
and
\begin{equation}
C(p_1,p_2)= \frac{W_{x_{1},x_{2}}(p_{1},p_{2})}{W_{x_{1},x_{2}}(p_{1})W_{x_{1},x_{2}}(p_{2})}=1.  \label{fullcoh}
\end{equation}

\subsection{Correlation femtoscopy in random phase representation}\label{CFRPR}
Both these cases of completely independent and fully coherent radiation from points $x_1$ and $x_2$, which correspond to mixed and pure quantum states,
can be reproduced in the formalism of partially coherent phases \cite{Tolst}. To demonstrate it,
let us consider the two-point source with coordinates $x_1$ and $x_2$ and some undetermined phases $\phi (x_i)$ ($i=1,2$) and  express amplitude of single boson emission with momentum $p$,
\begin{equation}
A_{x_1,x_2}(p)=e^{-iE\tau}\sum\limits_{i=1}^2 e^{ipx_i} e^{i \phi (x_i)}.
\label{superpos-random}
\end{equation}
The symmetrized amplitude of the two-boson radiation is
\begin{equation}
A_{x_1,x_2}(p_1,p_2)=A_{x_1,x_2}(p_1)A_{x_1,x_2}(p_2).
\end{equation}
The probability of registering the two identical particles with momenta $p_1$ and $p_2$ is
\begin{eqnarray}
\overline{W(p_1,p_2)}=\langle A_{x_1,x_2}(p_1)A_{x_1,x_2}(p_2)A_{x_1,x_2}^*(p_1)A_{x_1,x_2}^*(p_2)\rangle =   \nonumber \\
\frac{1}{4}\sum\limits_{i_1,i_2,i_1^{\prime},i_2^{\prime}=1}^2 e^{i(p_1x_{i_1}+p_2x_{i_2}-p_1x_{i_1^{\prime}}-p_2x_{i_2^{\prime}})}\langle e^{i (\phi (x_{i_1})+\phi (x_{i_2})-\phi (x_{i_1^{\prime}})-\phi (x_{i_2^{\prime}})}\rangle, 
\label{corr-random}
\end{eqnarray}
where brackets mean the averaging over all combinations of the two-boson radiation from the two points with probabilities $\rho_{i_1i_2}=\rho_{i_1}\rho_{i_2}=\frac{1}{4}$ ($\rho_{i}=\frac{1}{2}$ as in previous subsection) and over all events with mutually different phases in separate points. 
The random emission corresponds to the two-point phase average: 
\begin{equation}
\left\langle e^{i (\phi (x)-\phi (x^{\prime}))}\right\rangle =\delta_K^4(x-x^{\prime}),
\label{delta_K}
\end{equation}
 where $\delta_K$ is the Kronecker delta, $x=x_{i_1}\;\text{or}\;x_{i_2}$, $x^{\prime}=x_{i^{\prime}_1}\;\text{or}\;x_{i^{\prime}_2}$, and we omit here the discrete index $i$:  $x_{i_1} \rightarrow x_1$, $x_{i_1^{\prime}} \rightarrow  x_1^{\prime}$, etc., aiming to apply formalism to continuously distributed emitters.
The other two-point phase averages are zero. The four-point phase average in chaotic sources is expressed through the sum of the products of the two-point ones (\ref{delta_K}):
\begin{eqnarray}
&&\langle e^{i(\varphi(x_1)+\varphi(x_2)-\varphi(x_1^{\prime})-\varphi(x_2^{\prime}))}\rangle = \delta_K^4(x_1-x_1^{\prime})\delta_K^4(x_2-x_2^{\prime})+\nonumber \\ 
&& +\delta_K^4(x_1-x_2^{\prime})\delta_K^4(x_2-x_1^{\prime})-\delta_K^4(x_1-x_1^{\prime})\delta_K^4(x_1-x_2^{\prime})\delta_K^4(x_2-x_1^{\prime}).
\label{random-phases}
\end{eqnarray}
The last subtracted term eliminates double counting in the four-point phase average in (\ref{corr-random}) and (\ref{random-phases}), when $x_{i_1}=x_{i_2}=x_{i_1^{\prime}}=x_{i_2^{\prime}}$ and is usually omitted at the large number of emitters, but for essentially small number it is important. It corresponds to considered earlier elimination of the double counting in the model of independent emitters (cf. (\ref{wrong}) vs (\ref{correct})) when the two bosons are emitted from the same point. With accounting for the subtracted term in (\ref{random-phases}), the correct results (\ref{correct}) in the formalism of random phases follow directly from (\ref{corr-random}). 

If the source emission is not independent and fully coherent, it means that $\left\langle e^{i (\phi (x)-\phi (x^{\prime}))}\right\rangle =1$, and Eq. (\ref{corr-random}) leads to the result (\ref{fullcoh}). 

In the case of independent emitters the  phase average (\ref{delta_K}) can be presented in the form \cite{Tolst}:
\begin{eqnarray}
G_{x^{\prime}x}=\left\langle e^{i (\phi (x)-\phi (x^{\prime}))}\right\rangle =\delta^4(x-x^{\prime})=I_{x^{\prime}x}\delta(t-t^{\prime}),
\label{phase-aver}
\end{eqnarray}
where $I_{x^{\prime}x}$ is the overlap integral,
\begin{equation}
I_{x^{\prime}x}=\left|\int d^3{\bf r}\psi_{x}(\tau,{\bf r})\psi_{x^{\prime}}^*(\tau,{\bf r})\right|/N,
\label{Gij}
\end{equation}
where $N$ is the normalization of the wave function. As is demonstrated, the two-particle emission with fully random phases describes the  mixed state $\{i_1i_2\}$ corresponding to different combinations of the two-particle emission from points $x_{i_1}$ and $x_{i_2}$ with probabilities   $\rho_{i_1i_2}=\rho_{i_1}\rho_{i_2}$. Such a description is possible only if both bosons are emitted {\it independently}  from different points $x_i$ in  {\it distinguishable/orthogonal}  quantum  states \footnote{For indistinguishable quantum states one cannot define/measure the classical probability of the separate state in the system. Note also that the orthogonality  is the necessary but not sufficient condition for the independence of emission. }. The latter requirement is satisfied in the above case of the flat momentum spectrum for each emitter, $f(p)=W(p)=const$, since the initial quantum states $~\delta^3({\bf r}-{\bf x_i})$ taken in different points $i$  are orthogonal.

\subsection{Uncertainty principle and formalism of partially coherent phases \label{UncPr}}
If the momentum spectrum $f(p)$ is essentially not flat, this corresponds to the wave packets characterized by their centers $x$ and some finite width. In this realistic case one cannot consider quantum states with very close distances between emitter centers as distinguishable and independent because similar to the situation when two identical bosons are  emitted from the same point, such a system gives no contribution to the Bose-Einstein correlation function \cite{Lyub}, like a fully coherent state (\ref{fullcoh}). One can discriminate between the  different states with emission centers $x$ and $x^{\prime}$ only if they are approximately orthogonal: the overlap integral (\ref{Gij}) is small, $I_{x^{\prime}x} \ll 1$. In other words,  the  distance  between the centers of the emitters has to be larger than the width of  the  emitted  wave  packets. Since the latter is the inverse of the variance $\Delta p$ of the momentum spectrum, so $({\bf x-x^{\prime}})^2 > 1/\Delta p^2$. The latter expresses the uncertainty principle: one can discriminate the wave packet with center $x$ {\it without noticeable violation of the particle spectrum} if the measurement that localizes the particle's position somewhere inside the sphere with the center $x$ and the diameter not less than $1/\Delta p$, unambiguously points to the quantum state with center $x$, but not $x^{\prime}$. So, the distance between the emitter centers $\Delta x = \left|{\bf x-x^{\prime}}\right|$ should satisfy the uncertainty principle in the form  $\Delta x^2 \Delta p^2 > 1$. Then the states are almost distinguishable and orthogonal, so the radiation from both emitters can be considered as independent and well approximated within the random phase approach with probabilities $\rho_{i}$ for separate quantum states. 

In relativistic physics there is another uncertainty principle: the measurement of the particle's momentum $p$ has accuracy depending on the duration of the measurement $\delta t$, $\delta p  \sim 1/\delta t$ \cite{Lifshits}\footnote{In fact,  Ref. \cite{Lifshits} presents one of the forms of the uncertainty  principle for energy-time measurement.}. One can measure the time of particle emission without noticeable violation of the momentum spectrum with accuracy not better than $1/\Delta p$ and, then, the $\delta$-function in Eq. (\ref{phase-aver}) has to be smeared when one deals with wave packets. So, for normalized wave packets, one cannot use the random phase approximation if the distance in space and time between emitters is less than the width $1/\Delta p$ of the wave packet (in units $\hbar=c=1$). For  example, if $f(\textbf{p})=\frac{1}{(2 \pi
\Delta p^2)^{3/2}}e^{-\frac{\textbf{p}^2}{2\Delta p^2}}$, then $I_{x^{\prime}x}=e^{-\frac{\Delta p^2(\textbf{x} - \textbf{x}^{\prime})^2}{2}}$ at $t=t^{\prime}$, and
$\delta (t-t^{\prime})\Longrightarrow G_{x^{\prime}x}^t \rightarrow e^{-\frac{\Delta p^2(t - t^{\prime})^2}{2}}$. The last term expresses the uncertainty principle for energy-momentum \& time measurements.

Therefore, the fully random phases, corresponding to the standard femtoscopy approach \cite{Kopylov, Coc}, are possible only under some conditions which we have discussed: the phase average for quantum states emitted from the points $x$ and $x^{\prime}$, written in the form of Eqs. (\ref{phase-aver}), (\ref{Gij}),
\begin{eqnarray}
G_{x^{\prime}x}=\left\langle e^{i (\phi (x)-\phi (x^{\prime}))}\right\rangle =I_{x^{\prime}x}e^{-\frac{\Delta p^2(t - t^{\prime})^2}{2}},
\label{phase-aver-gen}
\end{eqnarray}
has to be close to zero. Here $I_{x^{\prime}x}$ is the overlap integral (\ref{Gij}), and $\Delta p$ is a variance of the momentum spectrum of emitters.

In the opposite case, when $(x-x^{\prime})^2\ll 1/\Delta p^2$, the states are indistinguishable and overlap integral (\ref{phase-aver-gen}) $G_{x^{\prime}x}\approx 1$ at $t \approx t^{\prime}$. Then the result looks like the one for fully coherent radiation (\ref{fullcoh}), which takes place for very close emitters \cite{Lyub}. So, in both limiting cases of chaotic and fully coherent emission Eq. (\ref{phase-aver-gen}) leads to physically obvious results, and therefore the quantity (\ref{phase-aver-gen}), $G_{x^{\prime}x}=I_{x^{\prime}x}e^{-\frac{\Delta p^2(t - t^{\prime})^2}{2}}$, is the natural  measure of distinguishability/indistinguishability and mutual coherence of the two emitted states caused by the uncertainty principle at any distance between the emission centers $(x^{\prime}_{\mu}-x_{\mu})$.  In this way we solve the problem of distinguishability of the quantum states $\psi_i$ associated with different emission points $x_i$: we always sum  $i$-amplitudes, not $i$-probabilities, and  average the resulting distributions/spectra over partially coherent phases in correspondence with the requirement of maximal possible  distinguishability and the independence of different $i$-states compatible with the uncertainty principle for momentum \& position and energy-momentum \& time  measurements.

\subsection{Femtoscopic correlations in multiparticle systems}

We discussed above the simplest situation with only two emitted bosons (pions). Typically, the basic formalism of correlation interferometry in multipion systems for inclusive momentum spectra is similar \cite{Heinz}: one can use the same basic formula (\ref{femtobasic}) with the following substitutions:  $W(p) \rightarrow p^0 \frac{d^3 N}{d^3 p}$, $W(p_1,p_2) \rightarrow p^0_1p^0_2\frac{d^6N}{d^3p_1d^3p_2}$, $\rho(x) \rightarrow S(x,p\equiv \frac{p_1+p_2}{2})$.  The correlation between the space-time position $x$ of the emission point and momentum $p$ of emitted particle, reflected in $p$-dependence of the emission function $S(x,p)$, leads to dependence of the interferometry radii in formula (\ref{Gauss}) on the mean momentum $p$ of pion pair, $R_i\rightarrow R_i(p)$ \cite{MakSin}. The $x-p$ correlation appears due to fast expansion of the multiparticle systems created in high energy $A+A$ (or even $p+p$) collisions.  The radii $R_i(p)$ are associated then with homogeneity lengths in $i$-direction in inhomogeneous expanding systems \cite{Sin}. So, one can carry the two-pion correlation results from the two-pion system to multiparticle systems if one applies the above substitutions. In fact, our quantum-mechanical  consideration is related to the rest frame of the homogeneous emitting subsystem forming the spectra in a vicinity of some momentum $p$.

Another aspect of multiparticle emission is that the two-pion wave function may not always be factorized out as for independent subsystems: $\Psi=\left| \pi^+\pi^+\right\rangle \left|X\right\rangle$, where state $X$ is related to the residual part of the total system.  The symmetrization/antisymmetrization procedure has to be applied to the total system wave function  $\Psi$. However, it can be difficult to provide it if one supposes the distinguishability of the radiation points as in the standard correlation interferometry method of independent sources \cite{Gold, Kopylov}.  For example, let us suppose that one pion with momentum $p_1$, $\pi^+(p_1)$ is the primary particle emitted from the fireball near the point $x_1$, and another pion $\pi^+(p_2)$ with the momentum $p_2$ is emitted together with particle $X$ at the resonance decay, $Res\rightarrow \pi^+(p_2) + X(p_3)$ (say, $\rho^0\rightarrow \pi^+(p_2) + \pi^-(p_3)$)  near the point $x_2$. Such a system may not have  $\pi(p_1)\leftrightarrow \pi(p_2)$  symmetry because

(i) if at the $\pi(p_1)\leftrightarrow \pi(p_2)$ exchange the momentum $p_3$ of the particle $X$ is preserved, then the momentum of emitted resonance is changed and gets the other value $p_1+p_3$. This new resonance state has typically another probability to form as compared to the previous resonance state. (In extreme cases it can even happen that the value $(p_1+p_3)^2$ instead of $p_{Res}^2=(p_2+p_3)^2$ excludes the resonance decay into the pair $\pi^+(p_1)$, $X(p_3)$ because of the kinematics). In general, all that suppresses  the interference effect \cite{Ledn2};

(ii) if the exchange of $\pi^+$ momenta is accompanied by the change of $p_3$ by the value $q=p_1-p_2$, $p_3 \rightarrow p^{\prime}_3=p_3+q $,  then the two configurations $\left| \pi_{x_1}^+(p_1)\pi_{x_2}^+(p_2)X(p_3)\right\rangle$ and $\left| \pi_{x_1}^+(p_2)\pi_{x_2}^+(p_1)X(p^{\prime}_3)\right\rangle$ become distinguishable and so cannot interfere;

(iii)  if at $\pi(p_1)\leftrightarrow \pi(p_2)$ the particles $X(p_3)$ together with $\pi^+(p_2)$ are emitted by the resonance $Res$ from the fireball near point $x_1$ and primary pion $\pi^+(p_1)$ radiates from $x_2$, the requirement of distinguishability of the emission points may lead to principal distinguishability of the emission points $x_1$ and $x_2$ of the particle $X$ that excludes the pion interference effect.

So, the exact symmetry of the total system at $p_1 \leftrightarrow p_2$ exchange can be lost in similar situations, and the two configurations $\left| \pi_{x_1}^+(p_1)\pi_{x_2}^+(p_2)\right\rangle$ and $\left| \pi_{x_1}^+(p_2)\pi_{x_2}^+(p_1)\right\rangle$ of the two-pion subsystem can interfere only partially. This has to be taken into account in the transport models used for high energy collisions.

The attempt to estimate  the limits of applicability of the standard correlation femtoscopy is done in Ref. \cite{Gol}. It is argued that when $\left| \pi_{x_1}^+(p_1)\pi_{x_2}^+(p_2)\right\rangle \leftrightarrow \left| \pi_{x_1}^+(p_2)\pi_{x_2}^+(p_1)\right\rangle$ then the classical phase-space position of the residual part of the system is changed by the value $q(x_1-x_2)$  along each direction $i$. If this change exceeds the size $2\pi\hbar$ of the elementary cell of the phase-space per one degree of freedom, then the {\it residual} system moves to another quantum state, and it can be, in principle, measured. Therefore the pion interference should disappear when $q_iR_i > 2\pi$ ($\hbar=c=1$), where $R_i$ is the effective system size/homogeneity length in $i$-direction. However, this argumentation fails already in the simplest case when the system contains only identical particles. Such a system can be symmetrized with distinguishable and independent radiation points, and there is no principal possibility to measure the exchange of the  momenta between two identical pions at any value of $q$ using for this aim the residual system containing the same sort of pions. We think that the real picture of the pion interference can be restored on the basis of  symmetrized/antisymmetrized amplitude of the {\it total} system with partially coherent phases in the way  described in subsection \ref{UncPr}.

\section{The correlation femtoscopy in Gaussian approximation}

Here we apply the basic ideas  discussed in the previous section to construct the simple analytical model accounting for uncertainty principle in the correlation femtoscopy. We use the non-relativistic approximation in the rest frame of the source which has Gaussian sizes corresponding to the homogeneity lengths $R_i$ in the corresponding part of the total expanding system. The transformation of the results to a global reference frame, where the source moves with 4-velocity $u^{\mu}$, depends on the concrete model used.

\subsection{Analytical model for fixed emitters \label{AnMod}}
Let us introduce the quantum state $\psi_{x_i}(p,\tau)$ corresponding to a boson with mass $m$ emitted at the time $t_i$ from the effectively finite space region with the center ${\bf x}_i$ as a wave packet with momentum dispersion $\Delta p$, and then propagating freely:
\begin{equation}
\psi_{x_i}(p,\tau)=e^{ipx_i -iE\tau}e^{i\varphi(x_i)}\tilde{f}(\textbf{p}),
\label{psi2}
\end{equation}
where $\varphi(x)$ is some phase and $\tilde{f}$ defines the primary momentum spectrum $f(\textbf{p})$ that we take in the Gaussian form with variance $\Delta p =k$:
\begin{equation}
f(\textbf{p})=\tilde{f}^2(\textbf{p})=\frac{1}{(2 \pi
k^2)^{3/2}}e^{-\frac{\textbf{p}^2}{2k^2}}.
\label{f}
\end{equation}

The amplitude of the single-particle radiation from some 4-volume at very large times $t_{\infty}$ can be written as a superposition of the wave functions $\psi_{x_i}(p)$ with some coefficients/distribution  $\rho(x_i)$ for emission 4-points $x_i$. In the continuous limit of a very large number of emitters we can omit discrete index $i$, and in the momentum representation the superposition looks like: 
\begin{equation}
A(p,\tau)=c\int d^4x \psi_{x}(p,\tau)
\rho(x),
\label{A}
\end{equation}
where $c$ is the normalization constant.
Let us select the two directions: parallel to $z$-axis (for systems that are created in collision processes it is the beam axis), which is marked by the index ``L'' and orthogonal to it, the transverse axis ``T''. The distribution $\rho(x)$ of the emission centers  is supposed to be the Gaussian one, so that the quantity $\rho^2({\bf x},t)$, 
being the probability distribution in the case of random phases $\phi(x_i)$, is normalized to unity: 
\begin{equation}
\rho(x)=\frac{1}{2\pi R_T\sqrt{R_L T}}e^{-\frac{{\bf x}_T^2}{4R_T^2}-\frac{x_L^2}{4R_L^2}-\frac{t^2}{4T^2}}.
\label{rho}
\end{equation}
The single-particle momentum distribution averaged over events with different phase distributions is
\begin{equation}
\overline{W(p)}=c^2 \int d^4x d^4x^{\prime}
e^{ip(x-x^{\prime})}\rho(x)
\rho(x^{\prime})\langle e^{i(\varphi(x)-\varphi(x^{\prime})} \rangle f(\textbf{p}).
\label{W}
\end{equation}
To calculate the phase average one needs first to calculate the overlap integral (\ref{Gij}). In non-relativistic approach the wave function of the  particle emitted at the moment $t$ from the  point $\textbf{x}$ in coordinate representation  is at some time $\tau$
\begin{equation}
\psi_{x}(\tau,{\bf r})=\frac{1}{(2\pi)^{3/2}}\int \widetilde{f}(\textbf{p})e^{i {\bf p (r-x})}e^{-i\frac{{\bf p}^2}{2m}(\tau-t)}d^3p.
\label{x-repres}
\end{equation}
Then the modulus of the overlap integral (\ref{Gij}) is
\begin{eqnarray}
I_{x^{\prime}x}=\left|\int d^3{\bf r}\psi_{x}(\tau,{\bf r})\psi_{x^{\prime}}^*(\tau,{\bf r})\right|=\frac{e^{-\frac{k^2 ({\bf x} - {\bf x^{\prime}})^2}{2(1+
k^4 (t-t^{\prime})^2/m^2)}}}{(1+k^4 (t-t^{\prime})^2/m^2)^{3/4}}.
\label{Gij1}
\end{eqnarray}

To provide calculations in analytical form, we substitute the squared time difference $(t-~t^{\prime})^2$ by the constant proportional to its mean value over emission region
\begin{equation}
(t-t^{\prime})^2 \rightarrow a \langle (t-t^{\prime})^2 \rangle = a \int d^4xd^4x^{\prime} \rho(x)\rho(x^{\prime}) (t-t^{\prime})^2 =4 a T^2\equiv \alpha T^2.
\end{equation}
As we will demonstrate later, the results obtained within such a prescription reproduce with  good accuracy the exact numerical calculations for momentum spectra and correlation functions with the values of $\alpha$ depending on the basic  parameters of emission.
Then the overlap integral (\ref{phase-aver-gen}) that accounts for the uncertainty principle takes the form
\begin{equation}
G_{xx^{\prime}}=\langle e^{i(\varphi(x)-\varphi(x^{\prime}))} \rangle = \frac{e^{-\frac{k^2 ({\bf x} - {\bf x}^{\prime})^2}{2(1+
\alpha k^4T^2/m^2)}}}{(1+\alpha k^4 T^2/m^2)^{3/4}}e^{-k^2(t-t^{\prime})^2/2}.
\label{appr-Gij1}
\end{equation}

Now one can obtain the one-particle spectrum (\ref{W}). It is presented below for the case when homogeneity lengths are equal, $R_T=R_L=R$:
\begin{equation}
\overline{W(p)}=
Ne^{-\frac{{\bf p}^2}{2 k^2}-\frac{2 {\bf p}^2 R^2}{1+4 k_0^2 R^2}-\frac{{\bf p}^4 T^2}{2 m^2 \left(1+4 k^2 T^2\right)}},
\label{W-result}
\end{equation}
where
\begin{equation}
k_0^2=k^2/(1+ \alpha k^4 T^2/m^2).
\label{k0}
\end{equation}

The two-particle spectrum averaged over events with partially coherent phases is
\begin{eqnarray}
\overline{W(p_1,p_2)}& = & c^4\int d^4x_1 d^4 x_2 d^4 x_1^{\prime} d^4 x_2^{\prime}e^{i(p_1
x_1+ p_2 x_2 -p_1 x_1^{\prime} - p_2x_2^{\prime})} f(\textbf{p}_1) f(\textbf{p}_2)
\rho(x_1)\rho(x_2)\rho(x_1^{\prime})\rho(x_2^{\prime}) \cdot \nonumber\\
& & \cdot \langle
e^{i(\varphi(x_1)+\varphi(x_2)-\varphi(x_1^{\prime})-\varphi(x_2^{\prime}))}\rangle.
\label{WW}
\end{eqnarray}
The 4-point phase correlator supposing emitters to be chaotic and independent in a maximum possible way permitted by uncertainty principle is decomposed into the sum of  products of the two-point correlators (\ref{phase-aver-gen}) and contains the three terms
\begin{equation}\label{eq39}
\langle
e^{i(\varphi(x_1)+\varphi(x_2)-\varphi(x_1^{\prime})-\varphi(x_2^{\prime}))}\rangle = G_{x_1x_1^{\prime}}G_{x_2x_2^{\prime}}+G_{x_1x_2^{\prime}}G_{x_2x_1^{\prime}}-G_{x_1x_2}G_{x_1x_2^{\prime}}G_{x_2x_1^{\prime}},
\end{equation}
where the third term removes the double accounting which appears in the sum of the first two terms when $x_1\approx x_2\approx x_1^{\prime}\approx x_2^{\prime}$. We emphasize that just the second term --- factor at crossing interference term (with cosine) --- has to be compensated for when all four points are very close to each other. The detailed discussion about the subtracted term is presented earlier in the subsection \ref{CFRPR}. Note that we use the product of the minimal number (three) of the two-point correlators in the subtracted term. The structure of (\ref{eq39}) coincides with one in Eq. (\ref{random-phases}) and is symmetric under simultaneous interchange $x_1\leftrightarrow x_2$ \& $x_1^{\prime}\leftrightarrow x_2^{\prime}$ as it is required by the $(p_1,p_2)$-symmetry of Eq. (\ref{WW}).  

Let us split transverse direction ``T'' into two: {\it out} which is directed along the total transverse momentum of the pair ${\bf p}_{1T}+{\bf p}_{2T}$ and {\it side}, which is orthogonal to the {\it out}-direction as well as to the longitudinal ``L'' axis.  Then the calculations of the one- and two-particle spectra lead to the following correlation function in the variables of the half of the bosonic pair momenta sum ${\bf p}=\frac{{\bf p}_{1}+{\bf p}_{2}}{2}$ and the momenta difference ${\bf q=p_1-p_2}$

\begin{equation}
C({\bf p},{\bf q})=\frac{\overline{W(p_1,p_2)}}{\overline{W(p_1)}\overline{W(p_2)}}=1+e^{- q_T^2R_T^2 \frac{4k_0^2R_T^2}{1+4k_0^2 R_T^2}-q_L^2R_L^2 \frac{4k_0^2R_L^2}{1+4k_0^2 R_L^2}-\frac{({\bf q}\cdot {\bf p})^2T^2}{m^2} \frac{4k^2T^2}{1+4k^2T^2}} - C_d({\bf p},{\bf q}) ,
\label{Cwos}
\end{equation}
where $\overline{W(p_i)}$ is defined by (\ref{W}), $k_0$ is (\ref{k0}), and the subtracted function $C_d$ eliminates the double accounting at the averaging of the  4-points phase correlator, it is associated with the third term in Eq. (\ref{eq39}). This term reduces the intercept of the correlation function as  one can see below.

\begin{figure}
\center
\vspace{-0.06\textheight}
\includegraphics[height=0.4\textheight]{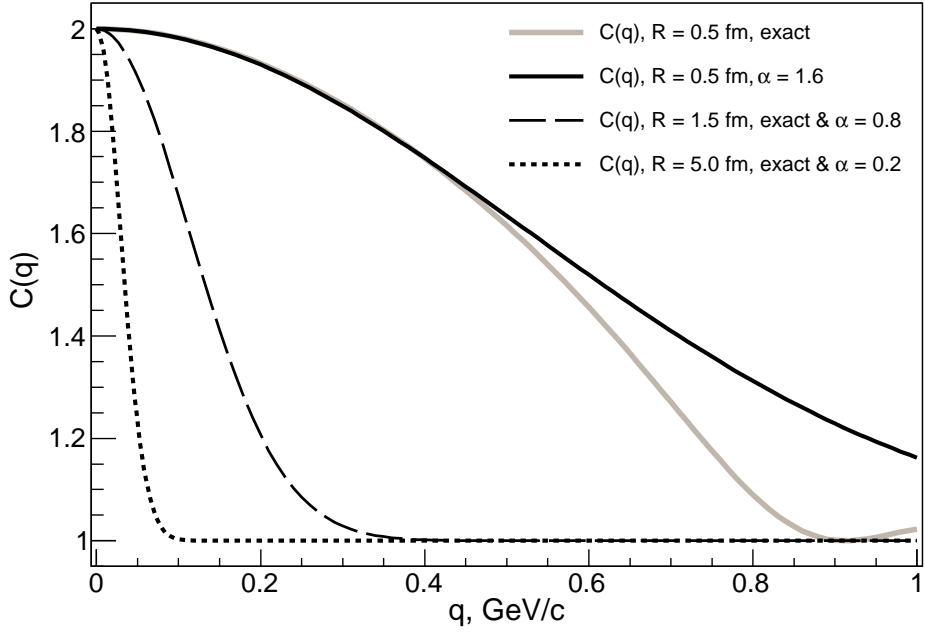}
\vspace{-0.02\textheight}
\caption{The comparison of the correlation functions in {\it side} direction without subtraction of the double accounting with corresponding analytical approximations. The $\alpha$ parameter values, which give a good agreement with the exact results, are presented for different system sizes. For $R$=1.5 and 5 fm the curves are merged, while for $R$=0.5 fm there is a disagreement with exact result at large $q$. The momentum dispersion $k=m=0.14$ GeV, $p$=0, T=R.}
\vspace{-0.01\textheight}
\label{fig0}
\end{figure}

The coefficient $\alpha$ in the expression for $k_0$  should be chosen from the requirement of a good agreement between the correlation function calculated using the approximation (\ref{appr-Gij1}) and the one calculated numerically using the exact expression for the space part of the phase correlator (\ref{Gij1}). For the typical freeze-out temperature $T_{f.o.}= m_{\pi}=m$ the value of $k$ is $k=\sqrt{T_{f.o.}m} = 0.14$ GeV. Having fixed the value of this parameter, we present in Fig.\ref{fig0} the comparison of the {\it side}-projections of the correlation functions calculated numerically using~(\ref{Gij1}) without subtraction of the double accounting and corresponding analytical expressions (the first two terms in (\ref{Cwos})). The values of $\alpha$ at different system sizes are presented. For fairly small systems, $R \approx  0.5 - 1.5$ fm, they are about unity, $\alpha \approx 1$, and are decreasing, $\alpha \rightarrow 0$, when  the system size $R$  grows up.

As follows from Eq. (\ref{Cwos}), the observed Gaussian interferometry radii of the system are reduced as compared to the standard  results (we mark it by {\it st} index) for the interferometry radii for the Gaussian source. The reductions for {\it side-} ($S$), {\it out-} ($O$)   and  {\it long-} ($L$) interferometry radii in the LCMS system ($p_L=0$) are
\begin{eqnarray}
\frac{R^2_{S}}{R_{S,st}^2} & =& \frac{4k_0^2R_T^2}{1+4k_0^2 R_T^2} \nonumber \\
\frac{R^2_{O}}{R_{O,st}^2} & = & \left(R_T^2\frac{4k_0^2R_T^2}{1+4k_0^2 R_T^2}+T^2v^2_{out}\frac{4k^2T^2}{1+4k^2T^2}\right)/\left(R_T^2+T^2v^2_{out}\right)  \label{radii-ratios} \\
\frac{R^2_{L}}{R_{L,st}^2} & = &\frac{4k_0^2R_L^2}{1+4k_0^2 R_L^2} \nonumber
\end{eqnarray}
where $v_{out}=p_{out}/m \ll 1$.

By means of Eq. (\ref{WW}) one can calculate the subtracting correlation function $C_d$ corresponding to the third term in Eq. (\ref{eq39}). It brings the result

\begin{equation}
C_d({\bf p},{\bf q})  = F(k_0^2R^2,k^2T^2) e^{-\frac{2{\bf q}^2 k_0^2 R^4  \left(1+8 k_0^2 R^2\right)}{\left(1+4 k_0^2 R^2\right) \left(1+8 k_0^2R^2+8k_0^4R^4 \right)}-\frac{2 k^2 T^4 ({\bf p} \cdot {\bf q})^2 \left(1+8 k^2 T^2\right)}{m^2 \left(1+4 p^2 T^2\right) \left(1+8 k^2 T^2+8 k^4 T^4\right)}}, 
\label{Cd1}
\end{equation}

\begin{equation}
F(k_0^2R^2,k^2T^2)  = \left(\frac{k_0}{k}\right)^{3/2}\frac{(1+4 k^2 T^2)^{1/2} \left(1+4 k_0^2 R^2\right)^{3/2}}{(1+8 k^2 T^2+8 k^4 T^4)^{1/2} \left(1+8 k_0^2 R^2+8 k_0^4 R^4\right)^{3/2}}. 
\label{Cd2}
\end{equation}

One can note that the pre-exponential coefficient $F(k_0^2R^2,k^2T^2)$ tends to zero at large $k_0^2R^2$ and/or $k^2T^2$ and tends to unity when both of these quantities tend to zero. In the former case of large sizes and/or hard radiation the intercept of the correlation function tends to 2  and in the latter case, when the sizes of the source are very small and/or radiation is soft, the intercept tends to 1. For example, at T=0
\begin{equation}
k^2R^2 \gg 1, \quad C({\bf p},{\bf q})=1+e^{-{\bf q}^2R^2},
\end{equation}
which is associated with the standard results, and
\begin{equation}
k^2R^2\ll 1, \quad  C({\bf p},{\bf q})\approx 1.
\end{equation}
The last result corresponds to the indistinguishable positions of the emitters due to the uncertainty principle in the case of very small sources and/or soft particle radiation.

\begin{figure}
\center
\vspace{-0.06\textheight}
\includegraphics[height=0.4\textheight]{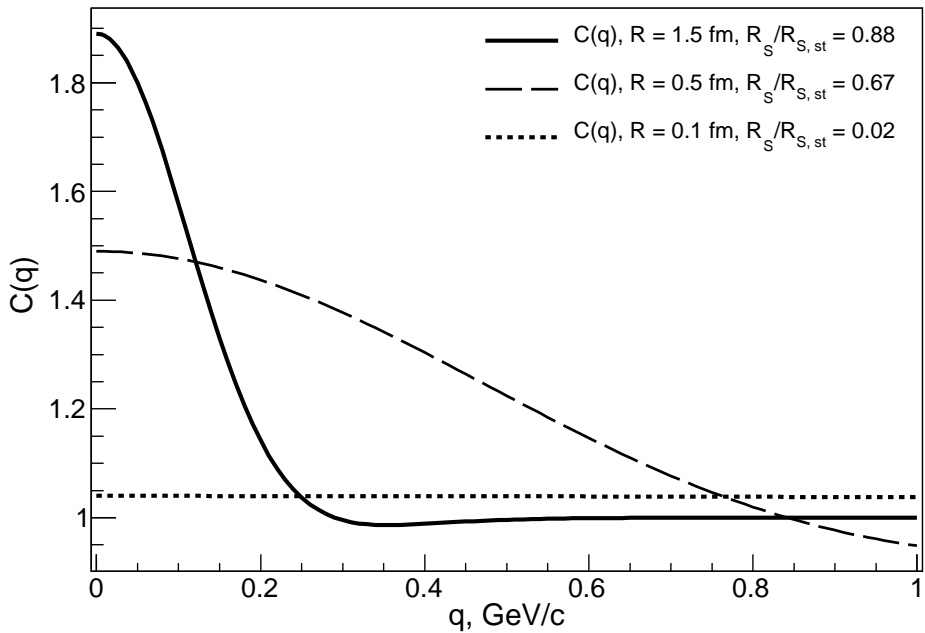}
\vspace{-0.02\textheight}
\caption{The behavior of the two-particle Bose-Einstein correlation function ({\it side}-projection), where the uncertainty principle and correction for double accounting are utilized. The momentum dispersion $k=m=0.14$ GeV, $p$=0, T=R.}
\vspace{-0.01\textheight}
\label{fig1}
\end{figure}

We illustrate realistic cases in Fig.~\ref{fig1} for $side$-projection of the correlation function.  Despite the fact that the correlation function at small $k^2R^2$ loses the Gaussian form presented by the second term in (\ref{Cwos}), the Gaussian fits still follow the tendencies of the analytic results (\ref{radii-ratios}), which demonstrate the reduction of the interferometry radii  when the sizes of the sources become small. The elimination of the double accounting also leads to the reduction of the intercept of the correlation function when the system size shrinks. So, there is  the positive correlation between observed interferometry radii and the intercept when the system size is changing. The analytic results reproduce the exact ones at $\alpha=2$ ($R=0.1$ fm), $\alpha=1.6$ ($R=0.5$ fm), $\alpha=0.8$ ($R=1.5$ fm).  It is worth noting that at large enough system sizes, $R > 2-3$ fm, the effects practically disappear. Note that such  marginal values are typical for the peripheral $A+A$ \cite{Ipp} and, probably, for $p+Pb$ collisions at the LHC. 

\subsection{Thermal density matrix in the wave packet formalism}

In the previous subsection we considered the analytical model where the particle emission events are localized within the sizes of the wave packets and are mutually independent and incoherent at the distances which exceed these sizes \footnote {Remember, that in the limiting case of the point-like emission from the 4-point $x_i$ the wave function of the single event in {\itshape momentum} representation is $\propto e^{ipx_i}$}. The important question is whether, if the system is thermal, this approach can reproduce the thermal distribution function in the thermodynamic limit of large globally equilibrated systems. The thermal density matrix for such systems is $\rho^{th} \propto e^{-H/T}$. For  ideal systems it is diagonal in the basis of  momentum eigenfunctions, which for one particle are the plane waves; in  the {\itshape coordinate} representation they are $\psi_p(x) \propto e^{ipx}$, where ($x=\tau, \textbf{r}$) and  $p$ is the particle 4-momentum. In the thermal system the plane waves obtain the weight $\rho^{th}_{pp} \propto e^{-{\bf p}^2/2mT}$, and the total density matrix for single particle states is in the coordinate representation
\begin{equation}
\rho(\textbf{r},\textbf{r}^\prime) \propto \int d^3p\:\rho_{pp}^{th}\:\psi_p(\textbf{r})\psi_{p}^{*}(\textbf{r}^{\prime})\propto e^{-mT(\textbf{r}-\textbf{r}^{\prime})^2/2}. 
\label{rho-tot}
\end{equation}
One can see that within the thermal de Broglie wavelength (without multiplier $\sqrt{2\pi}$) $\lambda_{coh}=1/\sqrt{mT}$  the density matrix has the non-diagonal terms. These terms in the density matrix have essentially quantum nature and express the interference effects and  phase correlations \cite{Balescu} within the distance $\lambda_{coh}$. Just the same area characterized by  $\lambda_{coh}$ with  $mT=\Delta p^2\equiv k^2$ is taken into account at the averaging of the phase correlations in this paper: at $t_1=t_2$ for Gaussian wave packets $\left\langle e^{i (\phi (x_1)-\phi (x_2))}\right\rangle =e^{-\frac{\Delta p^2(\textbf{x}_1 - \textbf{x}_2)^2}{2}}$. 

The Wigner function corresponding to the density matrix (\ref{rho-tot}) is 
\begin{equation}
f_W(p,r)=(2\pi)^{-3}\int d^3s\: e^{i\textbf{p}\textbf {s}}\rho(\textbf{r}-\frac{\textbf{s}}{2},\textbf{r}+\frac{\textbf{s}}{2})\propto e^{-{\bf p}^2/2mT}\rightarrow g(\textbf{r})e^{-{\bf p}^2/2mT}.
\label{wigner1}
\end{equation}

The Wigner function of the globally thermalized system does not depend on coordinates $r$; it means that the integral of $f_W(r,p)$ over $r$ is infinity. It reflects the fact that the plane waves are not localized wave functions. At the same time one can expect that for the systems with homogeneity lengths much larger than the mean particle wavelengths the finiteness correction is trivial and can be done in the way presented by the last  term in Eq. (\ref{wigner1}). Symbolically it is marked by arrow transition in Eq.  (\ref{wigner1}) to some normalized coordinate distribution function $g(r)$. Such a simple way is not correct, however, when the mean wavelength is comparable or less than the homogeneity lengths \footnote{For $g(r) \propto e^{-r^2/2R^2}$ the homogeneity length  coincides with the Gaussian radius $R$.} \cite{Sin-Spectr}. 
One possibility to find corrections for the system's finiteness is to change the plane wave basis --- the set of non-localized and coherent-over-whole-space wave functions --- to localized wave packets corresponding to a realistic case of local emissions from the different sites of the radiating (thermal) source. Below we present the example of an  essentially finite thermal model not pretending for a general theoretical consideration of the problem.
    
In subsection \ref{AnMod} as the basic functions are used the wave packets with the ``primeval'' momentum spectrum $f(p)\propto \exp (-p^2/2k^2)$ related to the single fixed emitter. One can take into account, however, that if one deals with a truly thermal system, then the velocities of the wave packets have to be thermalized, while until now we considered the fixed emitters which radiate the wave packets with zero mean momentum, ${\bf P} =\left\langle {\bf p}\right\rangle = 0$. Let us introduce non-zero mean momentum $\bf{P}$ for single wave packet \cite{P} and try to construct the finite system, which in the stationary case and thermodynamic limit tends to be described by the thermal Wigner function (\ref{wigner1}). To analyze the stationary system, one should consider the emission of the wave packets at any fixed time, say, $t=0$. The spectrum of the single emitter we choose in the thermal form is  $f(p,x_i)\propto e^{-(\textbf{p}-\textbf{P})^2/2mT}$. Then the wave function of such a single wave packet at $\tau=t=0$  takes the form:
 
\begin{itemize}
	\item in momentum representation
\begin{equation}
\psi_{x_i,P}(p,\tau=0)=e^{i(\textbf{p}-\textbf{P})\textbf{x}_i}e^{i\varphi(x_i,P)}e^{-(\textbf{p}-\textbf{P})^2/4mT},
\label{psip}
\end{equation}
	\item in coordinate representation
	\begin{equation}
	 \psi_{x_i,P}(r,\tau=0)= \frac{(2mT)^{3/4}}{\pi^{3/4}}\: e^{-mT ({\bf r}-{\bf x}_i)^2+i{\bf Pr}}e^{i\varphi(x_i,P)}.  
	\label{psix}
	\end{equation}

\end{itemize}
 One can note that the size of a single $i$-emitter is defined by the heat de Broglie wavelength $\sim 1/\sqrt{mT}$.

 In the wave packets' approach we consider the amplitudes, not probabilities, because of the quantum effects of interference and phase correlations within thermal de Broglie wavelength or, in other words, because of the uncertainty principle. So  we include the heat-like distribution $F({\bf P}) \propto \exp (-{\bf P}^2/2mT)$ for average momentum $P$ as a coefficient in the superposition of the wave packets with different $P$.  Then one can write, similarly to (\ref{A}), the full amplitude at $\tau=0$:
\begin{equation}
A(p)=c\int d^3xd^3P \rho({\bf x})F({\bf P}) \psi_{x,P}(p,\tau=0),
\label{A1}
\end{equation}
where the centers $x$ of the wave packet emission are distributed similarly to Eq. (\ref{rho}), where $\rho(x)=\frac{1}{(2\pi R^2)^{3/4}}e^{-\textbf{x}^2/4R^2}$ and quantity $\rho^2(x_i)$, which is the probability distribution of the emitter centers in the case of random phases $\phi(x_i)$, is normalized to unity.  
Let us suppose that
\begin{equation}
\left\langle e^{i (\phi (x_1,P_1)-\phi (x_2,P_2))}\right\rangle =\left\langle e^{i (\phi (x_1)-\phi (x_2))}\right\rangle\delta^3(P_1-P_2).
\label{rho2}
\end{equation}
Then the Wigner function is 
\begin{equation}
f_W(p,r)=(2\pi)^{-3}\int d^3s\: e^{-i\textbf{r}\textbf {s}}\left\langle A(\textbf{p}-\frac{\textbf{s}}{2})A^*(\textbf{p}+\frac{\textbf{s}}{2})\right\rangle \propto e^{-\frac{{\bf p}^2}{2mT}\frac{1+8mTR^2}{1.5+8mTR^2}}e^{-{\bf r}^2/2R^2}.
\label{wigner2}
\end{equation}

As one can see, the representation of the thermal system in the wave packets' basis that accounts for the coherence effects within heat de Broglie wavelength brings the standard results for the systems with homogeneity lengths larger than the thermal wavelength and different results for small enough systems. Noticeable deviations from the standard result take place for $mTR^2 < 1$. It is conditioned by the coherence effects between closely spaced emitters in the thermal system.  As for the femtoscopy scales, the detailed calculations demonstrate that noticeable deviations from the corresponding results of subsection \ref{AnMod} appear only for extremely small systems with effective sizes less than 0.5 fm at the typical temperatures T= 150 MeV.   
 
\section{Non-femtoscopic correlations}

The problem of non-femtoscopic correlations is of great importance for the interferometry analysis
of small systems and was studied in many papers, particularly in the recent paper~\cite{NF}. 
\textit{Non-femtoscopic} correlations are usually referred to as those not directly connected with
the spatiotemporal scales of the emitter, e.g., the correlations stemming from the momentum-energy conservation \cite{Chaj}. 
In this sense they differ from the QS and FSI correlations, which serve as the basis for the correlation femtoscopy method and are therefore often called \textit{femtoscopic} ones.

The strong interest in the question of the non-femtoscopy correlations is motivated in particular by the fact
that for relatively small systems they appreciably affect the complete two-particle correlation function, forming the so-called correlation baseline. It has an influence on the interpretation of the interferometry radii momentum dependence in $p + p$ collisions with high multiplicities \cite{Alice1,Alice2,Star}, where the different mechanisms of spectra formation are under discussion. Therefore, for successful and unambiguous application of the correlation femtoscopy method to the case of elementary particle collisions, one needs to know the mechanisms of non-femtoscopic correlations to separate the femtoscopic and non-femtoscopic ones. 
The important problem in this regard is whether we can factorize out the part corresponding to the non-femtoscopic correlations from the total correlation function. In the next two subsections we intend to investigate the possibility of such factorization with respect to the non-femtoscopy correlations of the two kinds having different physical origin: minijets fragmentation and event-by-event initial state fluctuation. The analysis is performed numerically within
the simple models of three- and two-particle emission accounting for the respective non-femtoscopic correlations.

\subsection{Non-femtoscopic correlations from minijets}

To trace in detail the genesis of non-femtoscopic correlations and their interplay with the femtoscopic ones
in our model, and also to make the model description more clear and consistent, we start our consideration
of three-particle emission from the case when emitted particles are totally uncorrelated, 
and then in a stepwise way include both types of correlations into the model. Thus, at first, QS correlations are assumed to be absent, as if the particles were considered as only \textit{equivalent} but  \textit{not identical} in the meaning common for quantum mechanics, and therefore distinguishable like in classics. Interactions between particles, and hence the FSI correlations, are also ignored.
 
In an such approach, if we deal with three particles having momenta $\textbf{p}_1$, $\textbf{p}_2$ and $\textbf{p}_3$ which are emitted independently with the amplitudes $A(p_1)$, $A(p_2)$ and $A(p_3)$, the three-particle emission probability is
\begin{equation}
\overline{W_0(p_1,p_2,p_3)}= \langle A(p_1) A^{*}(p_1)\rangle \langle A(p_2) A^{*}(p_2)\rangle \langle A(p_3) A^{*}(p_3)\rangle = W_0(p_1)W_0(p_2)W_0(p_3),
\label{w0}
\end{equation}
where the amplitude $A(p)$ has the form (\ref{A}) and averaging is performed over all random-phase emission events in the same way as in the subsection \ref{CFRPR}.

In the next step we switch on the non-femtoscopy correlations, supposing that two of three observed particles originate from a minijet, and the third one is emitted separately. Then to take into account the correlation between the former two particles, induced by the fact that they came from the same minijet (or cluster in momentum space), one should modify the expression~(\ref{w0}) for the three-particle emission probability \cite{NF} (factor $\Delta(p_1+p_2+p_3)$ will be explained later; at the moment it is just unity),
\begin{equation}
\overline{W_{NF}(p_1,p_2,p_3)}= \frac{1}{3!} \sum^3_{i\neq j=1} W_0(p_1)W_0(p_2)W_0(p_3)|Q(p_i, p_j)|^2\Delta(p_1+p_2+p_3).
\label{wcl3}
\end{equation}
Here the factor $ Q(p_i,p_j) = \exp{\left(-\frac{(p_i-p_j)^2}{2\gamma^2}\right )}$ describes mentioned ``minijet fragmentation'' correlation between the particles with momenta $\textbf{p}_i$ and $\textbf{p}_j$. The summation over
indices $i, j$ takes into account that cluster can be formed by different pairs of particles.

To also describe the correlations induced by the energy-momentum conservation law, one should include a delta-function like $\delta^4(p_1+p_2+p_3)$ into the right-hand side of Eq. (\ref{wcl3}). 
However, in our analysis we shall  use the ``soft'' form of the energy-momentum conservation law, where the delta function is substituted by its Gaussian-like approximation, keeping in mind that the conservation law being strictly kept for the whole system of produced particles can be only approximately fulfilled for the subsystem of identical bosons under consideration. Then the approximate momentum conservation takes the form 
\begin{equation}
\delta^4(p_1+p_2+p_3)\rightarrow \Delta(p_1+p_2+p_3)=Ce^{-(p_1+p_2+p_3)^2/d^2}.
\label{soft}
\end{equation}

The expressions for the single- and two-particle emission probabilities $W_{NF}(p_1)$ and $W_{NF}(p_1,p_2)$
can be found as the simple average of different $\overline{W_{NF}(p_1,p_2,p_3)}$ integrals over ``extra'' momenta, which correspond to different possible variants of which exact particle (particle pair) is observed. We have three such integrals in the single particle spectrum and six integrals in the two-particle one:
\begin{eqnarray}
W_{NF}(p_1)&=&\frac{1}{3}\sum_{i}\int d^4p^*_1d^4p^*_2d^4p^*_3 \delta^4(p_1-p^*_i)\overline{W_{NF}(p^*_1,p^*_2,p^*_3)}, \label{cl1}\\
W_{NF}(p_1,p_2)&=&\frac{1}{6}\sum_{i \neq j}\int d^4p^*_1d^4p^*_2d^4p^*_3 \delta^4(p_1-p^*_i)\delta^4(p_2-p^*_j)\overline{W_{NF}(p^*_1,p^*_2,p^*_3)}. \label{cl2}
\end{eqnarray}
where $\delta$ means Dirac delta.

The Eq. (\ref{soft}) defines now $\Delta$ in Eq. (\ref{wcl3}). The corresponding correlation function $C_{NF}(p,q)$ being the ratio of two-particle spectrum~(\ref{cl2}) to the product of single-particle spectra~(\ref{cl1}) is
\begin{equation}
C_{NF}(p,q)=\frac{W_{NF}(p_1,p_2)}{W_{NF}(p_1)W_{NF}(p_2)}=\frac{2}{3}C^{mix}(p,q) +\frac{1}{3}C^{jet}(p,q),
\label{ccl}
\end{equation}
where $C^{mix}(p,q)$ denotes the contribution to the correlation function corresponding to the pairs formed by the ``independent'' particle and the particle from the cluster, while $C^{jet}(p,q)$ is another contribution related to the pairs where both particles belong to the cluster.

Now it is time to include Bose-Einstein correlations in our consideration. To do this we should start 
considering our particles as \textit{identical} ones. This means that calculating the three-particle emission probability, we cannot anymore just sum the single-particle emission probability products corresponding to different emission variants as we did before. Instead of this we should compose the symmetrized amplitude of such emission as the sum of the amplitudes corresponding to different specific ways of final state realization
\begin{equation}
A(p_1,p_2,p_3)= \frac{1}{\sqrt{3!}} \sum^3_{i\neq j=1} A(p_1)A(p_2)A(p_3)Q(p_i, p_j) = \frac{1}{\sqrt{3!}} (A_{p_1} + A_{p_2} + A_{p_3}),
\label{acl}
\end{equation}
where $A_{p_i}$ denotes the amplitude of the state when particle with momentum $\textbf{p}_i$ \textit{does not} come
from the cluster. The desired probability is given by
\begin{equation}
\overline{W(p_1,p_2,p_3)}= \Delta(p_1+p_2+p_3) \langle |A(p_1,p_2,p_3)|^2 \rangle.
\label{wfull} 
\end{equation}
However, as an estimate in the current study we suppose that amplitudes $A_{p_i}$ in (\ref{acl}) related to different
cluster compositions are almost orthogonal, so that the interference terms between them are negligible, and so the expression (\ref{wfull}) for $\overline{W(p_1,p_2,p_3)}$
contains only the following diagonal terms~\footnote{Such a suggestion can be justified if one assumes that at different $i$ the amplitudes $A_{p_i}$  are almost isolated
from each other  in momentum space.}:
\begin{equation}
\overline{W(p_1,p_2,p_3)}= \frac{1}{3!} \Delta(p_1+p_2+p_3) \sum^3_{i\neq j=1} \langle A(p_1) A^{*}(p_1) A(p_2) A^{*}(p_2) A(p_3) A^{*}(p_3)\rangle |Q(p_i, p_j)|^2.
\label{wfull2} 
\end{equation}
This to some extent corresponds to the quasiclassical approximation  used in most event generators.  

The average 
\begin{eqnarray}
&&\langle A(p_1) A^{*}(p_1) A(p_2) A^{*}(p_2) A(p_3) A^{*}(p_3)\rangle = \nonumber \\
&&=c^6 \int d^4x_1 d^4 x_2 d^4 x_3 d^4 x_1^{\prime} d^4 x_2^{\prime}
d^4 x_3^{\prime} e^{i(p_1 x_1+ p_2 x_2 + p_3 x_3 -p_1 x_1^{\prime} - p_2x_2^{\prime}- p_3x_3^{\prime})} f(\textbf{p}_1) f(\textbf{p}_2)
f(\textbf{p}_3) \cdot \nonumber \\
&&\cdot \rho(x_1)\rho(x_2)\rho(x_3)\rho(x_1^{\prime})\rho(x_2^{\prime})\rho(x_3^{\prime}) \langle e^{i(\varphi(x_1)+\varphi(x_2)+\varphi(x_3)-\varphi(x_1^{\prime})-\varphi(x_2^{\prime})-\varphi(x_3^{\prime}))}\rangle
\end{eqnarray}
in (\ref{wfull2}) contains the 6-point phase correlator $\langle e^{i(\varphi(x_1)+\varphi(x_2)+\varphi(x_3)-\varphi(x_1^{\prime})-\varphi(x_2^{\prime})-\varphi(x_3^{\prime}))}\rangle $ which can be decomposed into the sum of products of three two-point correlators (\ref{phase-aver-gen})
\begin{eqnarray}
&&\langle e^{i(\varphi(x_1)+\varphi(x_2)+\varphi(x_3)-\varphi(x_1^{\prime})-\varphi(x_2^{\prime})-\varphi(x_3^{\prime}))}\rangle =
G_{x_1x_1^{\prime}}G_{x_2x_2^{\prime}}G_{x_3x_3^{\prime}}+G_{x_1x_2^{\prime}}G_{x_2x_1^{\prime}}G_{x_3x_3^{\prime}}+ \nonumber \\
&&+G_{x_1x_3^{\prime}}G_{x_2x_2^{\prime}}G_{x_3x_1^{\prime}}+G_{x_1x_1^{\prime}}G_{x_2x_3^{\prime}}G_{x_3x_2^{\prime}}+
G_{x_1x_2^{\prime}}G_{x_2x_3^{\prime}}G_{x_3x_1^{\prime}}+G_{x_1x_3^{\prime}}G_{x_2x_1^{\prime}}G_{x_3x_2^{\prime}}.
\end{eqnarray}
Here, however, we do not include the terms removing the double counting to simplify further calculations.

The three-particle emission probability $\overline{W_{BE}(p_1,p_2,p_3)}$, corresponding to the case when \textit{only} QS correlations between particles are considered, can be written as
\begin{equation}
\overline{W_{BE}(p_1,p_2,p_3)}=\langle A(p_1) A^{*}(p_1) A(p_2) A^{*}(p_2) A(p_3) A^{*}(p_3)\rangle.
\label{wbe3}
\end{equation}

Now to investigate the possibility of separation of the femtoscopic correlations from non-femtoscopic ones,
we introduce the full correlation function $C(p,q)$, including both femtoscopic (Bose-Einstein) and non-femtoscopic (minijet and conservation law) correlations, and $C_{BE}(p,q)$ accounting only for QS correlations. The corresponding single- and two-particle emission probabilities $W(p_1)$, $W(p_1,p_2)$ and $W_{BE}(p_1)$, $W_{BE}(p_1,p_2)$ necessary for the calculation of the correlation functions are obtained from (\ref{wfull2}), (\ref{wbe3}) similarly to the $C_{NF}(p,q)$ case (see Eq. (\ref{cl1}) and (\ref{cl2})). And the CFs themselves naturally are
\begin{eqnarray}
C(p,q)&=&\frac{W(p_1,p_2)}{W(p_1)W(p_2)}, \label{cfull} \\
C_{BE}(p,q)&=&\frac{W_{BE}(p_1,p_2)}{W_{BE}(p_1)W_{BE}(p_2)}. \label{cbe}
\end{eqnarray}

\begin{figure}
\center
\vspace{-0.03\textheight}
\includegraphics[height=0.35\textheight]{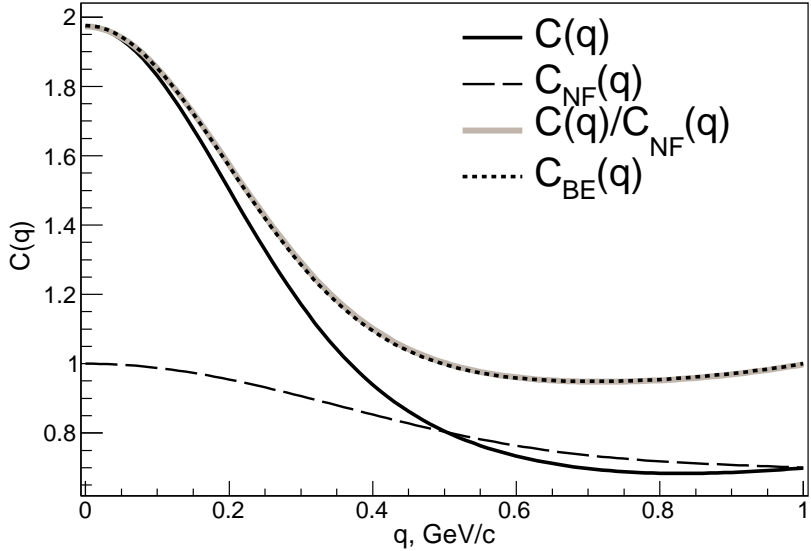}
\caption{The correlation function $C(q)$ (\ref{cfull}) accounting for the non-femtoscopic correlations due to minijet/cluster formation  and momentum conservation law, $C_{NF}(q)$ (\ref{ccl}), is compared with the pure QS correlation function $C_{BE}(q)$ (\ref{cbe}). One can see that the full CF is quite well factorized into the QS and non-femtoscopic parts. The model parameters are $\gamma = 0.5 $ GeV/\textit{c}, $p = 0.35$ GeV/\textit{c}, $k=0.1$ GeV/\textit{c}, $R=1 $ fm, $T=0$.}
\label{fig2}
\end{figure}

Figure \ref{fig2} illustrates the relation between the femtoscopic and non-femtoscopic correlations in our simple model. 
It appears that for a certain range of parameters consistent with more or less realistic description of the three-pion emission from a small system with the size~$\sim 1$~fm, $k \sim m_{\pi}$ and not very small $p$ the full correlation function can be roundly factorized in the two parts corresponding to the femtoscopic and non-femtoscopic correlations.

As the system size increases the accordance between $C_{BE}(q)$ and $C(q)/C_{NF}(q)$ gets worse, the both functions begin to differ by the value about a few percent. Such behavior at the system size increase can be explained by a  growing mismatch between the large system size and the small number (only three) of emitted particles.

\subsection{Non-femtoscopic correlations from fluctuating initial state}
Another type of correlations not induced by the quantum statistics effects are the correlations connected with existence of  subensembles of events with different emission functions that leads to the corresponding fluctuations in single-particle and two-particle momentum spectra. In hydrodynamical models of nucleus-nucleus and proton-proton collisions these fluctuations can be caused by asymmetrically fluctuating initial densities used for the hydro stage of the model.

Let us consider the effect of such correlations on the resulting correlation function in the example of a simple analytical model of two-particle emission.
Disregarding at first the QS correlations, we take into account the event-by-event emission function fluctuation by averaging the symmetrized two-particle emission probability over the ensemble of states corresponding to the events with different initial conditions
\begin{equation}
W_{NF}(p_1,p_2) = \Delta(p_1+p_2) \sum_{i} \rho(u_{i}) W_0(p_1; u_i) W_0(p_2; u_i),
\label{wnf}
\end{equation}
where $u_i$ denotes the $i^{th}$ type of initial conditions and $\rho(u_i)$ is the distribution over initial conditions, $\sum_i \rho(u_i) = 1$. The single-particle $u_i$-dependent probability $W_0(p_1;u_i)$ has the structure 
analogous to (\ref{W}), differing from the latter by the expression for the primary momentum spectrum
\begin{equation}
f(\textbf{p}) \propto e^{-\frac{\textbf{p}^2}{2k^2}}\rightarrow e^{-\frac{(\textbf{p}-\textbf{u}_i)^2}{2k^2}}. 
\label{fmod}
\end{equation}
The distribution $\rho(u_i)$ is supposed to be the Gaussian one, $\rho(\textbf{u}_i)= \frac{a^3}{\pi^{3/2}}e^{-\textbf{u}_i^2 a^2}$. 
As in the previous subsection, we use the soft form of momentum conservation law, expressed by the factor $\Delta(p_1+p_2)=C e^{-(p_1+p_2)^2/d^2}$.

To obtain averaged single-particle spectra $W_{NF}(p_1)$ and $W_{NF}(p_2)$ we integrate (\ref{wnf}) over corresponding
momenta:
\begin{equation}
W_{NF}(p_{1(2)})= \int d^4 p_{2(1)} W_{NF}(p_1,p_2).
\label{wnf1}
\end{equation}

The corresponding expression for the non-femtoscopic correlation function takes the form
\begin{equation}
C_{NF}(p_{1},p_{2})= \frac{W_{NF}(p_1,p_2)}{W_{NF}(p_1)W_{NF}(p_2)}.
\label{cbas}
\end{equation}

\begin{figure}
\center
\vspace{-0.02\textheight}
\includegraphics[height=0.35\textheight]{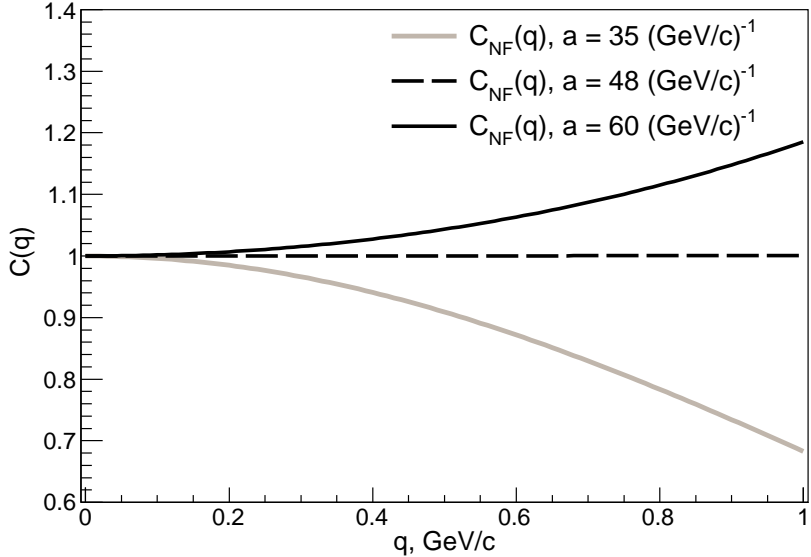}
\caption{The non-femtoscopic  correlation function (\ref{cbas}) (``baseline'') at the fluctuating initial conditions for hydrodynamic expansion of the system.
Depending on the relation between the value of the parameter $a$, characterizing the strength of the fluctuations, and the parameter $d = 1$ GeV/\textit{c}, which describes the softening of the momentum conservation law for the pion subsystem, the baselines can be growing, constant or decreasing with momentum difference $q$. The rest of the  parameters have the values $p = 0.35$ GeV/\textit{c}, $k=0.1$ GeV/\textit{c}, $R=1$ fm, $T=0$.}
\label{fig3}
\end{figure}

\begin{figure}
\center
\vspace{-0.02\textheight}
\includegraphics[height=0.35\textheight]{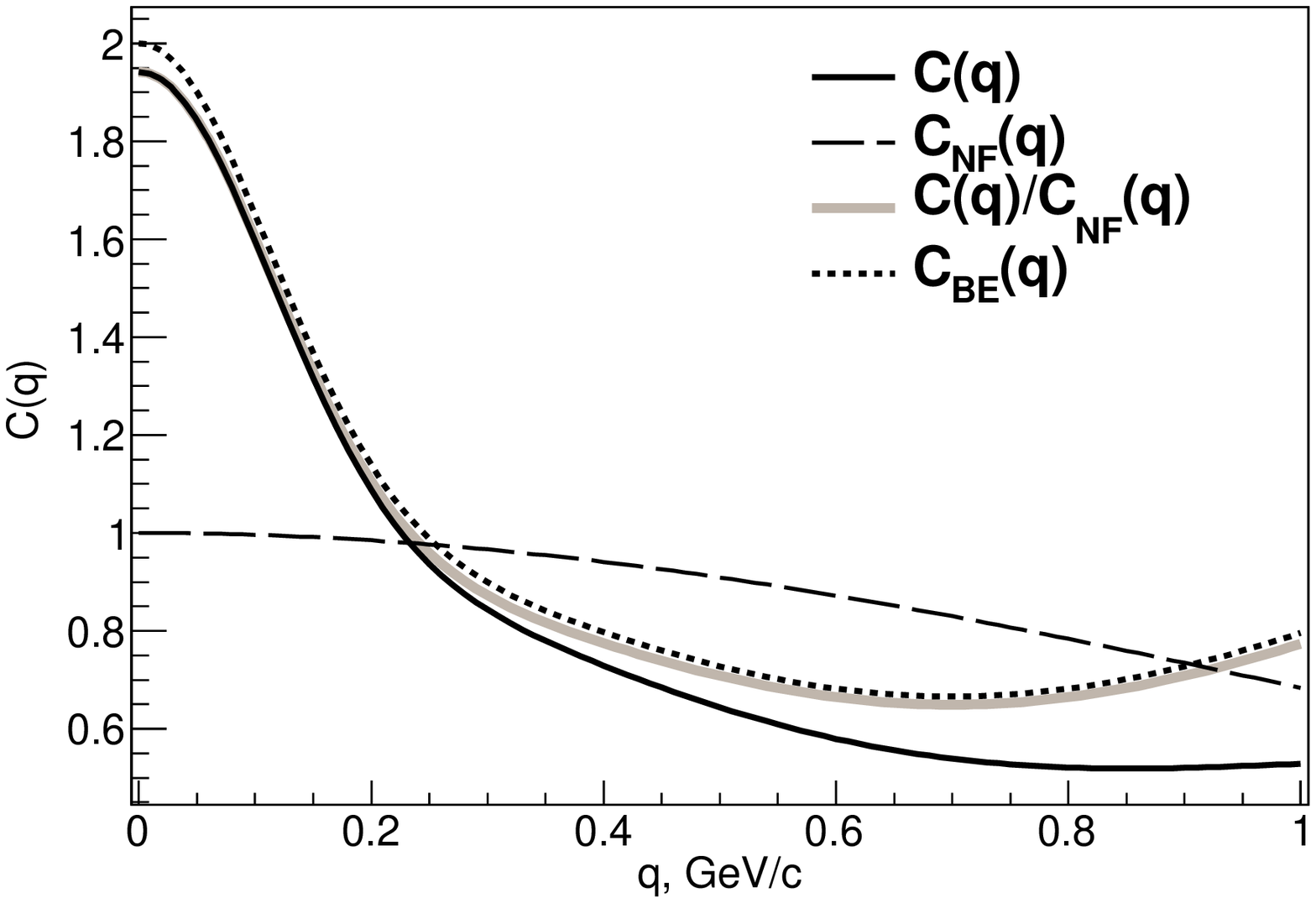}
\caption{Factorization of the femtoscopic and non-femtoscopic correlations when the latter are induced by fluctuating initial conditions for hydrodynamics. The parameters $a=35$ (GeV/$c)^{-1}$, $d = 1$ GeV/\textit{c}, $p = 0.35$ GeV/\textit{c}, $k=0.1$ GeV/\textit{c}, $R=1$ fm, $T=0$. }
\vspace{-0.005\textheight}
\label{fig4}
\end{figure}

Depending on the ratio between parameters $d$, describing momentum conservation, and $a$, describing initial conditions fluctuations, the baseline formed by the $C_{NF}$ function can be growing, constant or decreasing with $q$ (see Fig. \ref{fig3}).

The full correlation function $C$ is found similarly to the $C_{NF}$ case, starting with the appropriate expression for the averaged two-particle spectrum. To obtain the latter, one needs to take the Eq. (\ref{wnf}) and substitute the product $W_0(p_1; u_i) W_0(p_2; u_i)$ in it by the $u_i$-dependent two-particle spectrum accounting for the QS correlations. This one coincides with (\ref{WW}) up to the primary momentum spectrum $f(\textbf{p})$, which should be modified according to (\ref{fmod}).

The correlation function $C_{BE}$, accounting only for Bose-Einstein correlations, can be derived analogously
to $C$ and $C_{NF}$ ones taking $W(p_1,p_2)$ in the form (\ref{WW}), which can be considered as the limiting
case of (\ref{wnf}) when $\textbf{u}_{i}=0$ and $\Delta(p_1+p_2)$ is unity. The single-particle spectra $W(p_{1,2})$ are then calculated using (\ref{wnf1}). 

We see that fluctuating initial conditions for hydro can cause non-femtoscopic correlations similar to those resulting from minijet fragmentation. Figure \ref{fig4} illustrates the possibility of approximate factorization of its contribution to the full correlation function at the values of main model parameters introduced in the previous subsection and decreasing baseline.
Quite a similar situation takes place also for the case of growing baseline. However, when the baseline is constant, the 
three $C$, $C_{BE}$ and $C/C_{NF}$ curves just merge. Also, as opposed to the previously analyzed minijet correlations case, for the fluctuation correlation part of full correlation function the possibility to be factorized out is observed not only for small systems, but also, and even more pronounced, for the large ones.

\section{Conclusions}
We discussed the principal problems of the correlation interferometry of small sources, in particular, the similarity and the principal difference between Hanbury Brown \& Twiss intensity interferometry and pion correlation interferometry methods. The main subject of the paper is to build the  correlation femtoscopy
method that is going beyond the standard approach of independent/random particle emission.
It is found that the uncertainty principle leads to (partial)
indistinguishability of closely located emitters, which fundamentally
impedes their full independence and incoherence. The partial
coherence of emitted particles is because of the quantum nature of particle
emission and happens even if there is no specific mechanism
to produce a coherent component of the source radiation. The measure of distinguishability/indistinguishability and mutual coherence of the two emitted wave packets is associated with their overlap integral.  In thermal systems the role of corresponding coherent length is played by the thermal de Broglie wavelength. 

The formalism of partially coherent phases in the amplitudes of located close individual emitters is developed for the quantitative analysis. 
The specific treatment is required for elimination of the double accounting to the correlation function from the configurations in the four-point phase correlator when all individual emission centers are close. It is shown that the effects  are significant for the systems with the sizes and emission duration times about 1 fm, and they are expressed in the reduction of the interferometry radii and suppression of the Bose-Einstein correlation functions. There is the positive correlation between the source size and the intercept of the correlation function. At the typical slopes of the momentum spectra in $p+p$ and $A+A$ collisions these effects are negligible for fairly large sources with radii about or more than 2--3 fm.

The peculiarities of the non-femtoscopic correlations caused by minijets and fluctuations of the initial states of the systems formed in $pp$ and $e^+e^-$  collisions are analyzed also.  The factorization property for the contributions of femtoscopic and non-femtoscopic correlations into the complete correlation function is demonstrated in a wide range of parameter values in the simple analytic models.

The application of the theoretical results to $pp$ and $e^+e^-$ collisions will be done in separate works.

\section{Acknowledgments}
The authors are grateful to S. Akkelin, R. Lednicky and E. Sarkisyan-Grinbaum for fruitful discussions. The research was carried out within the scope of the EUREA: European Ultra Relativistic
Energies Agreement (European Research Group: ``Heavy ions at ultrarelativistic energies'') and is
supported by the National Academy of Sciences of Ukraine (Agreement 2013) and by the State
Fund for Fundamental Researches of Ukraine (Agreement 2013).

\end{document}